\documentclass[prl,twocolumn,superscriptaddress,letterpaper]{revtex4}
\usepackage{graphicx,amssymb,bm,natbib,amsfonts,amsmath}
\usepackage{hyperref}

\begin{document}

\title{Metal-insulator transition of the reduced surface of yttria-stabilized zirconia near Pt electrodes}

\author{D. A. Siegel}
\affiliation{Materials Physics Department, Sandia National Laboratories,
Livermore, CA 94550, USA}

\author{F. El Gabaly}
\affiliation{Materials Physics Department, Sandia National Laboratories,
Livermore, CA 94550, USA}

\author{K. F. McCarty}
\affiliation{Materials Physics Department, Sandia National Laboratories,
Livermore, CA 94550, USA}

\author{N. C. Bartelt}
\affiliation{Materials Physics Department, Sandia National Laboratories,
Livermore, CA 94550, USA}

\date{\today}

\begin{abstract}
The electrochemical reactions of solid oxide fuel cells occur in the region where gas-phase species, electrode, and electrolyte coincide.  When the electrode is an ionic insulator and the electrolyte is an electronic insulator, this “triple phase boundary” is assumed to have atomic dimensions.  Here we use photoemission electron microscopy to show that the reduced surface of the electrolyte yttria-stabilized zirconia (YSZ) undergoes a metal-insulator transition near Pt negative electrodes.  YSZ’s electron conducting region functions as an extended triple phase boundary that can be many microns in size, depending on oxygen pressure, temperature, applied voltage, and time.
\end{abstract}

\maketitle

In many electrochemical systems the performance of a cell is fundamentally limited by the surface area of a triple phase boundary, the region where gas-phase species, mobile ions, and mobile electrons coincide\cite{Nowotny}.  When the electrode is an ionic insulator, ions cannot be transported through the electrode and electrons cannot be transported through the electrolyte, so the triple phase boundary (TPB) has a narrow, nearly one-dimensional, edge geometry at the junction between the electrode, electrolyte and gas phase.\cite{Imbihl}  Since the detailed properties of this boundary region can dominate the behavior of a fuel cell, a major goal in electrochemistry is to understand the nature of ionic incorporation and the length scale of this reactive phase boundary.\cite{Maier, Yan}

A fundamental question is what precisely determines the size of the TPB.  The size will depend sensitively on the electrical properties of the electrolyte’s surface near the electrode,\cite{Kasamatsu} so answering this question requires understanding how the local surface properties differ from the bulk electrolyte when polarized in oxidizing or reducing environments.  Here we examine this issue for the prototypical solid oxide electrolyte\cite{deSouza} yttria-stabilized zirconia (YSZ) by examining its reduced surface near a Pt negative electrode.  

There have been some proposals in the literature that the electronic conductivity in the TPB might be enhanced compared to bulk YSZ. Previous work by Casselton suggested that the YSZ in this region might behave like a `virtual' electrode,\cite{Casselton} acting like a mixed conductive electrode when a voltage is applied; while other works have shown electrical contributions to conductivity in nonstoichiometric YSZ.\cite{Levy,Schouler}  YSZ has been shown to exhibit bulk `blackening'\cite{Casselton}, and a reduction zone has been shown to develop near the interface with a platinum electrode\cite{JanekInstability, JanekInsitu, Luerssen}.  However, the electronic properties of the TPB are difficult to study directly: a local probe\cite{Kumar} is required to selectively measure the region near the interface without altering the electrical properties of the system; probing the electronic structure of materials is nontrivial in general; and an electric potential and an oxygen-containing gas must be present in order for the cell oxygen side to be probed under operating conditions.  Photoemission electron microscopy (PEEM) is an ideal tool for addressing these issues.  PEEM with an imaging energy analyzer is capable of measuring the electronic density of states as a function of position on the sample, in real time and space, using photoelectrons as a non-contact probe.\cite{Zhang}

Using PEEM, we directly demonstrate the formation of a mixed conductive phase: electrochemical reduction causes the YSZ surface to undergo a transition from electronic insulator to conductor, with a finite electronic density of states at the Fermi level.  We observe a sharp phase boundary between insulating and conducting regions, and determine that this effect is localized to the near surface of the YSZ crystal.  We measure that the geometrical length of this conductive virtual electrode depends sensitively on the time, oxygen pressure, applied voltage, and temperature of the functioning electrochemical cell.

The substrates in this study are commercially grown and polished single-crystal fully stabilized YSZ(100) wafers from MTI Corp. (8 mole \% Y$_2$O$_3$).  The measurement temperature is 900K unless indicated otherwise, high enough that YSZ becomes an ionic conductor and no charging effects were observed.  PEEM is performed with an Elmitec SPELEEM with VG Scienta helium UV source, with 200 meV resolution. 

\begin{figure}
\includegraphics[width=8.5 cm] {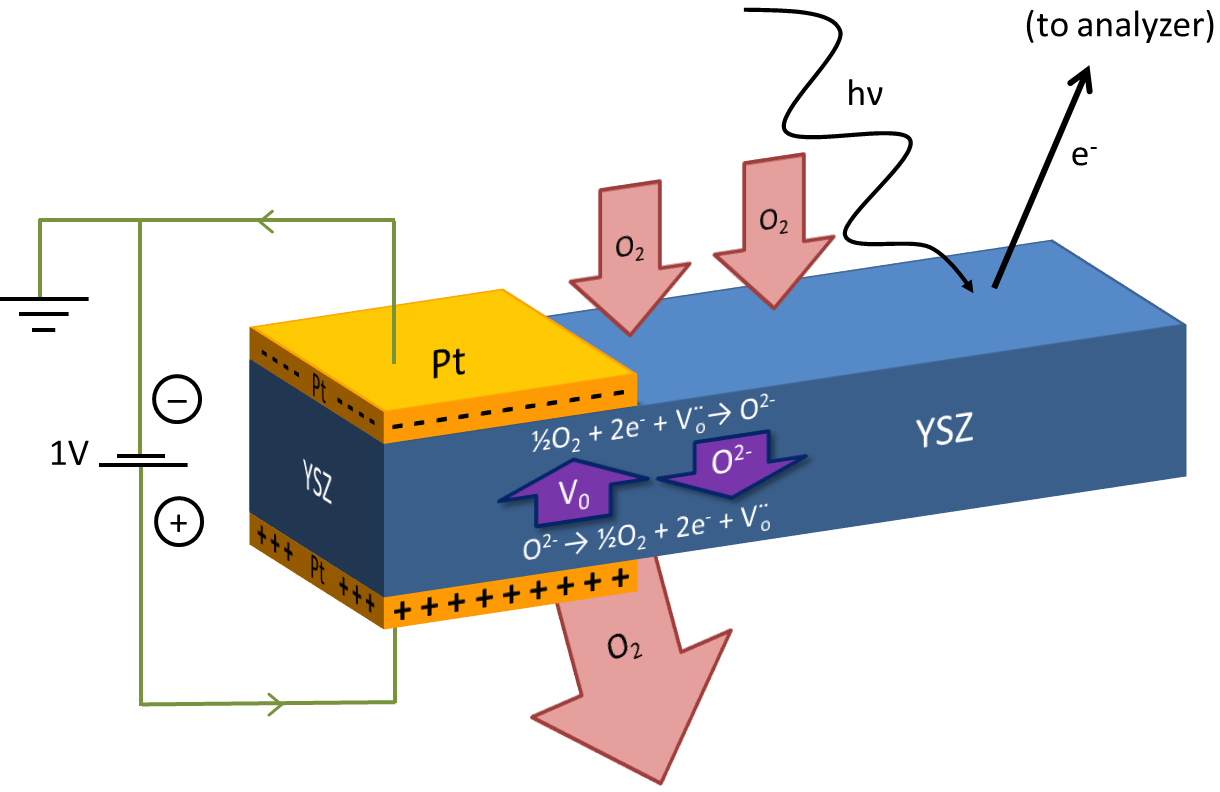}
\caption{(Color online) Schematic of the experiment.  PEEM is used to examine the YSZ surface near the negative Pt electrode. }
\end{figure}

We use the model Pt/YSZ/Pt electrolyzer cell shown in Figure 1, which allows us to emulate the oxygen side of a real fuel cell without the need of separate environments for H$_2$ and O$_2$.\cite{ElGabaly}  Both electrodes are exposed to the same oxygen environment.  An applied bias triggers the oxygen incorporation on the negative electrode and oxygen evolution on the positive electrode.

Figure 2 shows PEEM images from YSZ adjacent to a Pt electrode, where the Pt is the bright vertical stripe on the left side. The PEEM contrast of the YSZ near the electrode changes as a voltage is applied to the cell, starting from the negative electrode and extending tens of microns from the electrode-electrolyte interface.  The YSZ develops a sharp boundary between two types of PEEM contrast. The boundary moves to the right with time.  In past experiments, an increase in total (energy-integrated) PEEM intensity in the growing phase was attributed to a reduced form of YSZ with a smaller work function.\cite{JanekInstability} 

\begin{figure}
\includegraphics[width=8.5 cm] {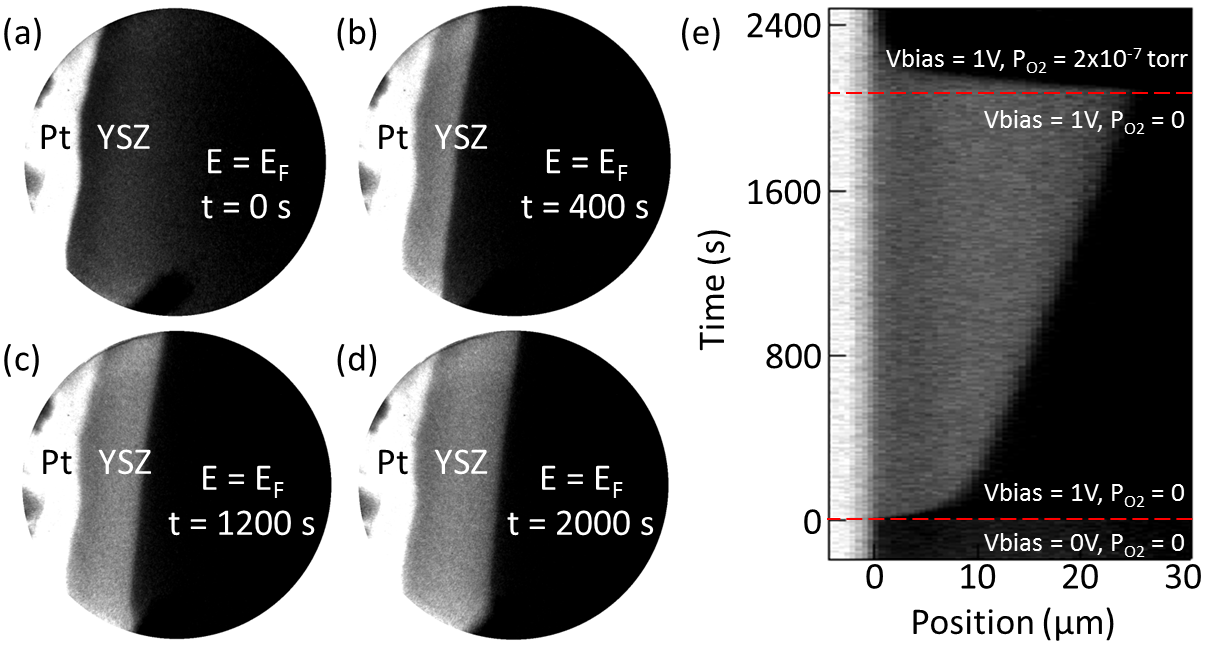}
\caption{(Color online)  Time evolution of a YSZ surface after a voltage is applied to the electrodes.  (a-d) Before t=0, the sample sits in ultra-high vacuum (UHV) with no voltage applied.  At t=0, 1 V is applied to the electrodes.  The YSZ develops two PEEM contrasts: a dark gray, and a medium gray that grows from the platinum electrode (light gray).  Field of view is 75 $\mu$m.  (e) An intensity profile as a function of time.  The medium gray YSZ appears to increase indefinitely in response to an applied voltage (t=0), but shrinks in response to 2$\times$10$^{-7}$ torr of oxygen gas (introduced at t=2100). }
\end{figure}

In our experiment, the PEEM images are produced with an imagining energy analyzer.  For instance, Fig. 2 uses only photoelectrons from the Fermi energy (E$_F$) for imaging.  Therefore the intensity in Fig. 2 represents the angle-integrated density of states at the Fermi level.  Wherever the contrast in Fig. 2 is bright, the surface has a high density of states at E$_F$, and where the contrast in Fig. 2 is dim, the density of states at E$_F$ is low.  Therefore, platinum appears relatively bright, as expected, and the electronically insulating YSZ is dimmer due to its bandgap ($\sim$5 eV\cite{PaiVerneker, Paje, Vohrer}). But since the YSZ has \textit{two} shades of gray, the two regions of YSZ must have differing densities of states at the Fermi level.

\begin{figure}
\includegraphics[width=8.5 cm] {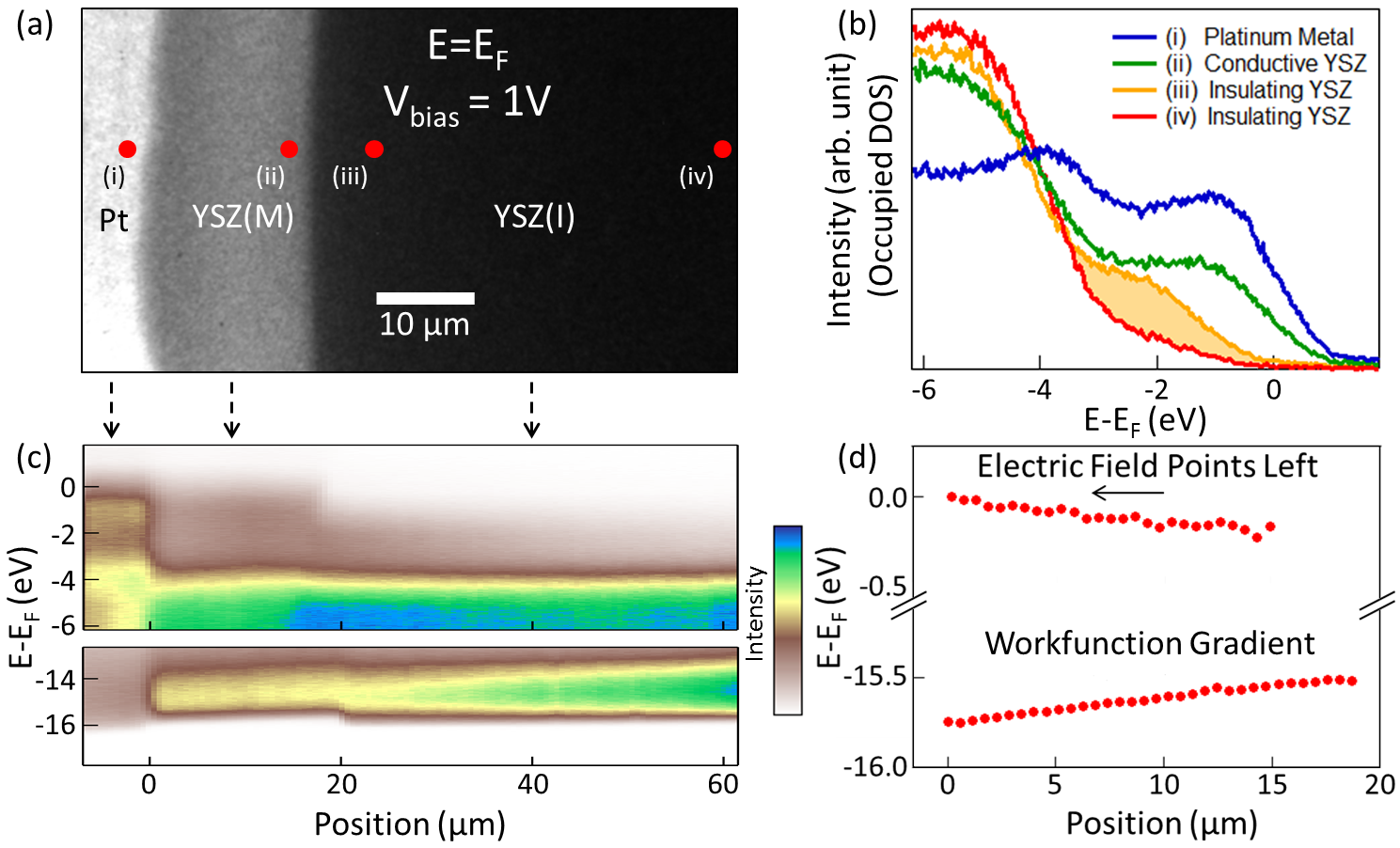}
\caption{(Color online)  PEEM Density of States (DOS) (a) PEEM Image of the Pt/YSZ surface at the Fermi energy.  Pt and YSZ(M) have a significant density of states at the Fermi level and are bright in this image.  (b) DOS extracted from various positions in panel a (labelled i-iv) show the difference between the Pt, YSZ(M), and YSZ(I), and demonstrates the effect of increasing oxygen vacancy density.  (c) DOS as a function of position and binding energy.  The boundary between Pt and YSZ(M) is labelled as x=0, while the boundary between YSZ(M) and YSZ(I) occurs at x$\approx$20 μm (the size of the YSZ(M) increased slightly during the measurement).  The intensity in the top and bottom panels have been normalized separately, but use the same color scale.  (d) The position of the Fermi level and secondary cutoff in panel (b), showing that they have slopes of opposite signs due to gradients in the electric field and electron workfunction. }
\end{figure}

In Fig. 3 we investigate the electronic structure more thoroughly at the locations marked by red circles in Fig. 3(a). The density of states of the platinum electrode in Fig. 3(b) has a sharp Fermi cutoff with a finite density of states at the Fermi level, the expected features of a typical metal.  In the YSZ far from the electrode at position (iv), the red curve in Fig. 3(b) shows a $\sim$4 eV separation between the valence band and the Fermi level, with a vanishing density of states at the Fermi level.  Moving towards the electrode from position (iv) to position (iii), the density of states inside the bandgap increases gradually, denoted by the orange shaded region in Fig. 3(b).  However, stepping across the phase boundary from the YSZ(I) at position (iii) to the YSZ(M) at position (ii), the leading edge of the density of states suddenly jumps to lower energy, and the Fermi edges of the YSZ(M) and the platinum electrode become aligned.  The YSZ(M) is therefore a mixed(M) electronic-ionic conductor, or highly-doped n-type semiconductor,\cite{Maier, Schouler, Gellings} while the YSZ(I) is an electronic insulator(I), and their interface is an electronic metal-insulator boundary.  The sharp increase in intensity near the Fermi level at about $\sim$18 $\mu$m in Fig. 3c clearly shows the boundary.  Since the YSZ(M) occurs at the negative electrode, we associate it with a reduced form of YSZ with more oxygen vacancies than the YSZ(I).

In an oxygen ambient the net oxygen reduction reaction at the negative electrode takes the form:
\begin{equation}
\label{eq:eq1} \frac{1}{2}O_2+V^{\bullet\bullet}_o+2e^- \longrightarrow O_o
\end{equation}
where $V^{\bullet\bullet}_o$ is a positively charged oxygen vacancy, O$_o$ is a lattice oxygen in YSZ, and O$_2$ is a gas molecule that evolves from the positive electrode.  The important point is that this reaction can only take place in the presence of mobile electrons; the electronic conductivity of the YSZ(M) allows reaction (1) to take place entirely on the YSZ(M) without building up charge.

The key issue we next address is how the YSZ(M) varies as a function of position, time, and the cell's gas-phase environment.  To investigate the spatial dependence of the YSZ(M) phase, we measured the positions of the leading edges in PEEM at low and high binding energy, which allows us to extract the electric potential and the electronic component of the electrochemical potential.\cite{Vohrer}  The lower panel of Fig. 3d shows that the electron work function of the YSZ(M) has a gradient and changes by 0.2 eV along the surface, which is consistent with an oxygen vacancy gradient (the magnitude of the gradient varies inversely with the width of the YSZ(M); see Supplementary Information).  The lower panel of Fig. 3c shows that the workfunction decreases by 0.3 eV across the YSZ(M)/YSZ(I) interface, and that the workfunction of the YSZ(M) is higher than the YSZ(I).  This is different from past work\cite{JanekInstability, JanekInsitu}, where only the total (energy integrated) PEEM intensity was measured: energy-resolved imaging shows us that an overall increase in PEEM intensity may not be related to a change in the work function.

Figure 3(d) also shows that the Fermi edge is at lower binding energy close to the platinum electrode; the electric field therefore points along the surface towards the Pt electrode, with a magnitude that varies roughly inversely with the YSZ(M) width.  We also find that the change in electric potential across the metallic platinum electrode is negligible, as expected.

A crucial question is whether the conducting phase persists in a more oxidizing environment, allowing reaction (1) to continue on the YSZ.  Fig. 2(e) shows that when the oxygen pressure is low, the width of the YSZ(M) increases without limit.  But when the oxygen pressure is sufficiently high, it approaches a finite steady-state value.  Figure 4(a) reveals that the YSZ(M) phase can be many microns wide in an oxygen ambient, but narrows with increasing oxygen pressure.

For the contribution of YSZ(M) to the reaction (1) to be relevant at high pressures, the decrease of width with oxygen pressure cannot be too large: suppose the oxygen absorption rate varies proportional to (Width $\times$ Pressure), and the width varies like P$^{-x}$.  Then if x$\textless$1 the reactivity of YSZ(M) will increase with pressure.  In fact, the slope of the log-log plot in the inset of Fig. 4(a) gives x = 0.76$\pm$0.04, which suggests that the YSZ(M) can enhance the electrochemistry over a wide range of conditions even as the oxygen pressure continues to increase.  The pressure-dependent size of the TPB accounts for the measurements of Ref.\cite{Gur},  which show that the exposure of YSZ to O$_2$ decreases the activity of YSZ for NO decomposition.

Figure 4(b) shows that the width increases with applied voltage, a result consistent with $^{18}$O tracer experiments\cite{Opitz} and I-V measurements\cite{Rutman} that suggest increased triple phase boundary areas with voltage.  Figure 4(c) shows that the asymptotic length scale increases as a function of temperature, likely due to the increased reactivity and ionic mobility.  The detailed explanation for this behavior will be elaborated upon in a subsequent work where we will present a model of vacancy transport: in general, the length of the mixed conductive YSZ(M) depends on the interplay between oxygen adsorption and incorporation, evolution into the vapor phase (the reverse reaction), and diffusive transport along the surface.\cite{Siegel}

Figure 2(e) also shows that an oxygen dose of $\textless$30 Langmuir  is capable of converting a 20 $\mu$m wide region of YSZ(M) to YSZ(I).  We can thus establish an upper limit to the depth of this reduced surface layer (a lower limit is difficult to estimate): if we assume zero oxygen ion current, a $\geq$1\% change in oxygen concentration\cite{Levy}, and an oxygen sticking coefficient of $\leq$1, the reduced surface layer must be $\leq$900 nm deep, which is much smaller than the 20 $\mu$m surface width.  Therefore, we find that this mixed conductive phase is essentially a surface phenomenon, in contrast to past observations of bulk YSZ `blackening'.

\begin{figure*}
\includegraphics[width=14 cm] {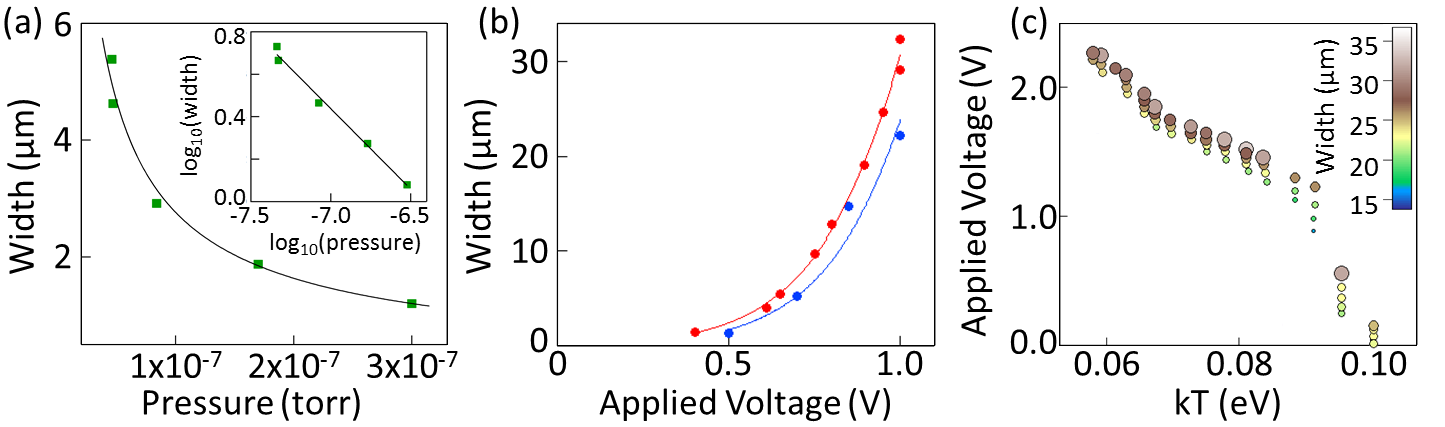}
\caption{(Color online)  Response of a virtual electrode to perturbations. (a) Width of the conductive YSZ as a function of oxygen pressure after long times (V = 1V, T = 900K).  The line is a guide to the eye of the form $L = \alpha P^{-0.76}$.  Inset: A log-log plot of the data gives a power law with a slope of -0.76$\pm$0.04.  (b) Width of the conductive YSZ as a function of applied voltage after long times (T = 900K, P = 4$\times$10$^{-9}$ torr).  Lines are guides to the eye; blue and red data are taken with decreasing and increasing applied voltage, respectively.  (c)  The width of the conductive YSZ as a function of applied voltage and temperature.  The size and color of the experimental data points correspond to the width of the conductive YSZ, taken with an oxygen pressure of 1.1$\times$10$^{-6}$ torr.  }
\end{figure*}

The reactivity of the YSZ(M) increases the active area for oxygen adsorption from a nearly 1-dimensional triple phase boundary to a 2-dimensional region in the general vicinity of the electrodes. The platinum electrode is essential, since the  phase conversion initiates adjacent to it, but our results suggest that the platinum electrode need not be involved in the oxygen ion adsorption, reduction, or transport. 

In conclusion, we directly observe that the solid oxide electrolyte YSZ exhibits a sharp transition from electronic insulator to electronic conductor under an applied voltage adjacent to a negative, reducing Pt electrode.  While more than one mechanism of oxygen incorporation into YSZ may exist in this system, these direct experimental observations demonstrate the creation of an extended triple phase boundary.  This improved understanding of the mechanisms of oxygen incorporation can help inform the design of electrode geometries and the operation conditions of electrochemical cells.\cite{Yan, Kumar, OpitzGeometry, Ryll} 

\begin{acknowledgments}
We would like to thank Ivan Ermanoski and Taisuke Ohta for useful conversations.
Our research was supported by the Office of Basic Energy Sciences, Division of Materials and Engineering Sciences, U.S. Department of Energy under contract no. DE-AC04-94AL85000
\end{acknowledgments}

Correspondence and requests for materials should be addressed to DavidASiegel@gmail.com or Bartelt@Sandia.gov.

\end{document}


\title{Supplemental Information: Metal-insulator transition of the reduced surface of yttria-stabilized zirconia near Pt electrodes}

\author{D. A. Siegel}
\affiliation{Materials Physics Department, Sandia National Laboratories,
Livermore, CA 94550, USA}

\author{F. El Gabaly}
\affiliation{Materials Physics Department, Sandia National Laboratories,
Livermore, CA 94550, USA}

\author{K. F. McCarty}
\affiliation{Materials Physics Department, Sandia National Laboratories,
Livermore, CA 94550, USA}

\author{N. C. Bartelt}
\affiliation{Materials Physics Department, Sandia National Laboratories,
Livermore, CA 94550, USA}

\date{\today}

\maketitle

In our primary manuscript we discuss the behavior of the work function and Fermi level when a voltage is applied across the electrodes, and state that they both vary roughly inversely with the size of the YSZ(M) region.  To demonstrate this, we present Figure S1, which has data from the low kinetic energy (workfunction) cutoff and the high kinetic energy (Fermi level) cutoff of the YSZ(M) for three different measurements.

What we observe in Fig. S1 is that the slopes of both cutoffs vary with the size of the YSZ(M) region, even when the voltage across the electrodes remains constant.  For the Fermi level, this indicates that the total drop in electric potential across the YSZ(M) the overpotential remains roughly constant in time, and that the magnitude of the electric field decreases with increasing YSZ(M).  Similar behavior in the work function data may be the result of Arrhenius behavior for the vacancy concentration.

The work function data in Fig. S1(a), specifically curves (iv) and (v), are roughly linear in the center of the YSZ(M) but curve up and down at the ends.  This may be an effect of the finite lateral resolution of our measurement.  In Fig. S1b we have plotted the slope of a linear fit to the entire curves (iv) and (v) in black, while the slopes of just the central segments of those curves are plotted in grey.

\begin{figure}
\includegraphics[width=8.5 cm] {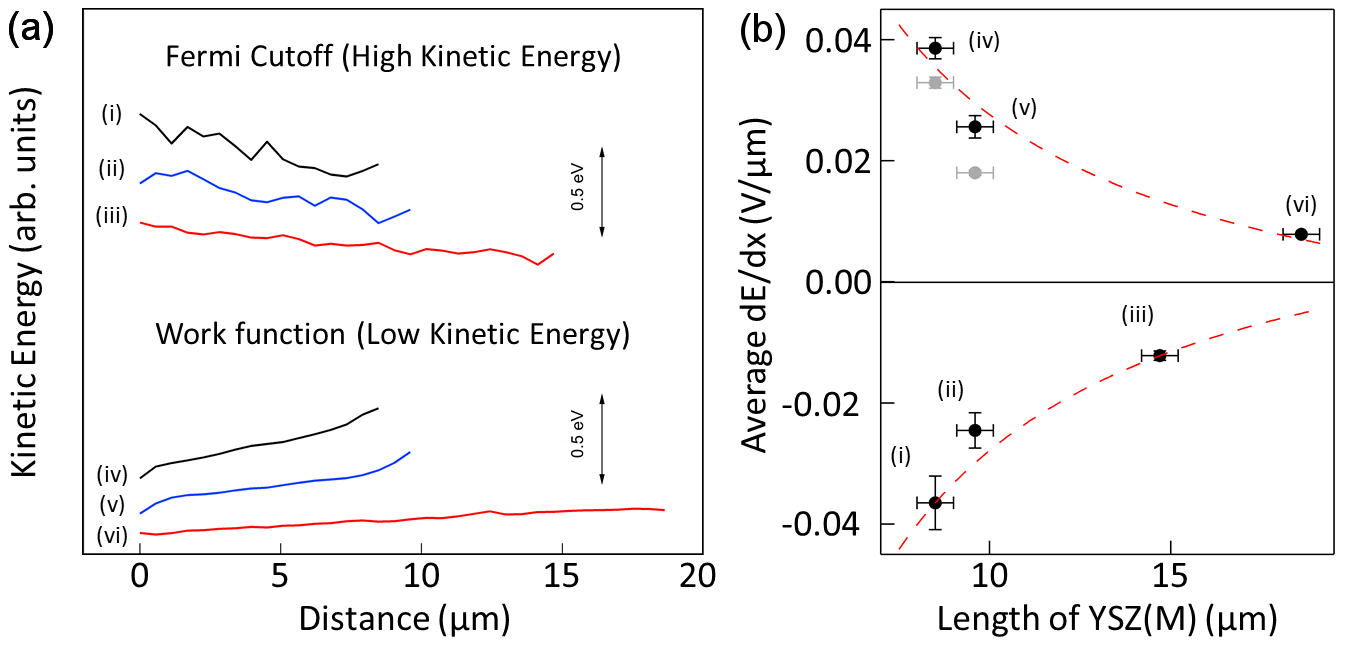}
\caption{(Color online) (a) The energies of the Fermi level and workfunction cutoffs are plotted for three different measurements at 1 V, 900 K, and UHV conditions, labelled (i)-(vi).  The data has been given arbitrary offsets for ease of viewing, so 0.5 eV scale bars are given on the right side of the image.  (b) The average slopes of these datasets are given in black.  Two data points are given in light grey to indicate the slopes of the tangents at the center of lines (iv) and (v).  The red dashed lines are guides to the eye of the form y = a/x+b.
}
\end{figure}